\documentclass[a4paper,12pt]{article}
\pdfoutput=1
\usepackage{graphicx, rotating,amssymb}
\ifx\pdfoutput\undefined
\usepackage[dvips,bookmarks]{hyperref}	% This is for arXiv.org
\else
\usepackage{hyperref}	% This is for pdftex
\fi
\hypersetup{colorlinks,bookmarksopen,bookmarksnumbered,citecolor=verdes,
linkcolor=blus,pdfstartview=FitH,urlcolor=rossos}

\def\hhref#1{\href{http://arxiv.org/abs/#1}{#1}} % in bibliography
\usepackage[margin=0pt,font=small,labelfont=bf]{caption}
\usepackage{amsmath}
\usepackage{multicol}
\usepackage{color}
\definecolor{rosso}{cmyk}{0,1,1,0.4}
\definecolor{rossos}{cmyk}{0,1,1,0.55}
\definecolor{rossoc}{cmyk}{0,1,1,0.2}
\definecolor{blu}{cmyk}{1,1,0,0.3}
\definecolor{blus}{cmyk}{1,1,0,0.6}
\definecolor{bluc}{cmyk}{1,1,0,0.1}
\definecolor{verde}{cmyk}{0.92,0,0.59,0.25}
\definecolor{verdec}{cmyk}{0.92,0,0.59,0.15}
\definecolor{verdes}{cmyk}{0.92,0,0.59,0.4}

\font\tenrsfs=rsfs10 at 12pt
\font\sevenrsfs=rsfs7
\font\fiversfs=rsfs5
\newfam\rsfsfam
\textfont\rsfsfam=\tenrsfs
\scriptfont\rsfsfam=\sevenrsfs
\scriptscriptfont\rsfsfam=\fiversfs
\def\mathscr#1{{\fam\rsfsfam\relax#1}}

\newcommand{\fig}[1]{~\ref{fig:#1}}
\newcommand{\eq}[1]{~{\rm (\ref{eq:#1})}}

\def\circa#1{\,\raise.3ex\hbox{$#1$\kern-.75em\lower1ex\hbox{$\sim$}}\,}

\newcommand{\beq}{\begin{equation}}
\newcommand{\eeq}{\end{equation}}

\def\circa#1{\,\raise.3ex\hbox{$#1$\kern-.75em\lower1ex\hbox{$\sim$}}\,}
\makeatletter

\def\art{\@ifnextchar[{\eart}{\oart}}
\def\eart[#1]#2#3#4#5#6{{\rm #2}, {#3 #4} {\rm (#6) #5} [{\hhref{#1}}]}
\def\hepart[#1]#2{{\rm #2, \hhref{#1}}}
\newcommand{\oart}[5]{{\rm #1}, {#2 #3} {\rm (#5) #4}}

\newcounter{alphaequation}[equation]
\def\thealphaequation{\theequation\hbox to
0.6em{\hfil\alph{alphaequation}\hfil}}
\def\eqnsystem#1{
\def\@eqnnum{{\rm (\thealphaequation)}}
\def\@@eqncr{\let\@tempa\relax \ifcase\@eqcnt \def\@tempa{& & &} \or
  \def\@tempa{& &}\or \def\@tempa{&}\fi\@tempa
  \if@eqnsw\@eqnnum\refstepcounter{alphaequation}\fi
\global\@eqnswtrue\global\@eqcnt=0\cr}
\refstepcounter{equation} \let\@currentlabel\theequation \def\@tempb{#1}
\ifx\@tempb\empty\else\label{#1}\fi
\refstepcounter{alphaequation}
\let\@currentlabel\thealphaequation
\global\@eqnswtrue\global\@eqcnt=0 \tabskip\@centering\let\\=\@eqncr
$$\halign to \displaywidth\bgroup \@eqnsel\hskip\@centering
$\displaystyle\tabskip\z@{##}$&\global\@eqcnt\@ne
\hskip2\arraycolsep\hfil${##}$\hfil& \global\@eqcnt\tw@\hskip2\arraycolsep
$\displaystyle\tabskip\z@{##}$\hfil
\tabskip\@centering&\llap{##}\tabskip\z@\cr}
\def\endeqnsystem{\@@eqncr\egroup$$\global\@ignoretrue} \makeatother

\begin{document}
\begin{center}
%{XXX-xx/2008-xx}
{ \hfill SACLAY--T09/027}
\color{black}

\vspace{1cm}

{\Huge\bf Anti-deuterons\\[4mm] from heavy Dark Matter}

\medskip
\bigskip\color{black}\vspace{0.6cm}

{
{\large\bf Carolin B. Br\"auninger}$^{a, b}$,
{\large\bf Marco Cirelli}$^b$
}
\\[7mm]
{\it $^a$ Institut f\"ür Theoretische Physik, Universit\"at T\"übingen,\\
Auf der Morgenstelle 14, D-72076 T\"ubingen, Germany
}\\[3mm]
{\it $^b$ Institut de Physique Th\'eorique, CNRS, URA 2306 \& CEA/Saclay,\\ 
	F-91191 Gif-sur-Yvette, France}\\%[3mm]
%{\it $^b$ Affiliation 2}\\[3mm]
%{\it $^c$ Affiliation 3}
\end{center}

\bigskip

\centerline{\large\bf Abstract}
\begin{quote}
\color{black}\large
We study the fluxes of anti-deuterons that could be produced by annihilations in the galactic halo of Dark Matter particles with multi-TeV mass and a large annihilation cross section, as indicated by the recent PAMELA results. The model of Minimal Dark Matter (MDM) is an example in this category.
We find that the fluxes are well within the reach of planned experiments for DM candidates that annihilate mainly into quark pairs, and also extend into the multi-GeV range above the background. They are instead suppressed and concentrated at too large energies to emerge above the background if the main annihilation channel is into gauge bosons (such as in particular $W^+W^-$ in the MDM case).
\end{quote}

%\tableofcontents
%\newpage

\section{Introduction}

Cosmology and astrophysics provide several convincing evidences of the existence of Dark Matter~\cite{reviews}. The observation that some mass is missing to explain the dynamics of galactic or cluster rotations dates back to the '30s~\cite{rotation}. The observations from weak lensing~\cite{lensing}, for instance in the spectacular case of the so-called `bullet cluster'~\cite{bullet}, provide evidence that there is mass where nothing is optically seen. More generally, global fits to a number of cosmological datasets (Cosmic Microwave Background, Large Scale Structure and also Type Ia Supernovae) allow to determine very precisely the amount of DM in the global energy-matter content of the Universe at $\Omega_{\rm DM} h^2 =0.110 \pm 0.005$~\cite{cosmoDM}.
 
All these signals pertain to the gravitational effects of Dark Matter of the cosmological and extragalactical scale. Searches for explicit manifestation of the DM particles that are supposed to constitute the halo of our own galaxy have instead so far been giving negative results, but this might be on the point of changing. 

Indeed, a recent flurry of results from a number of cosmic ray experiments has shown a few surprising anomalies that can be interpreted as the first indirect signatures of annihilations of DM particles in the galactic halo, on top of the standard cosmic rays (CR) of astrophysical origin.\footnote{It is entirely possible that these anomalies will be explained in terms of a better understanding of standard CR~\cite{Blasi} or rather peculiar astrophysics, such as one or more local pulsars~\cite{pulsars}, nearby sources of CR in galactic spiral arms~\cite{Piran} or exploding stars~\cite{exploding}. We are of course instead interested in the possibility that DM is at the origin of the signals.} We review below the precise indications of these experiments, but we anticipate that the central piece of our motivations are the PAMELA results, that indicate the possibility of a heavy DM particle (multi-TeV) that can annihilate with a large cross section (of the order of $10^{-23}\, {\rm cm}^3/{\rm sec}$ or more, much larger than the expected thermal cross section of~ $\approx 3\cdot 10^{-26}\, {\rm cm}^3/{\rm sec}$) into quarks or SM gauge bosons.

In indirect DM searches, the key point is to look for channels and ranges  of energy where it is possible to beat the background from ordinary astrophysical processes. This is for instance the basic reason why searches focus on fluxes of anti-particles, much less abundant in the Universe than the corresponding particles. 
The proposal to look in particular for fluxes of anti-deuterons in the cosmic rays was put forward for the first time in ref.~\cite{Donato1999}. The basic observation was that anti-deuterons are rarely produced in standard processes. Especially the low energy range (below about 1 GeV) is kinematically disfavored for astrophysical production and therefore open to the possible emergence of an exotic contribution like the one from DM. 
More recently, anti-deuterons from DM annihilations have been studied in~\cite{Profumo2005} and~\cite{Donato2008}. 

Motivated, as said, by the PAMELA results, we here focus our analysis on the case of DM particles that are heavy (multi-TeV masses) and have a large galactic annihilation cross section.
We compute and present the fluxes of anti-deuterons adopting a model independent approach, i.e. considering different possible masses and primary annihilation channels. But we also study the specific predictions of the model of Minimal Dark Matter~\cite{MDM}, that naturally falls into this class. 

The rest of the paper is organized as follows. In sec.~\ref{DM} we describe the DM models on which we concentrate our attention and their motivation on the basis of the data. In sec.~\ref{prodandprop} we describe the production of anti-deuteron from DM annihilations and its propagation in the galaxy, discussing in particular the astrophysical parameters and the associated uncertainties. In sec.~\ref{results} we present our results, comparing them with the expected sensitivities of planned experiments. In sec.~\ref{conclusions} we draw the conclusions.

\section{The Dark Matter framework}
\label{DM}

The recent results from the PAMELA satellite~\cite{PAMELA} show
\begin{itemize}
\item[$\cdot$] a significant excess above the expected smooth astrophysical background and a steep rise of the positron fraction $e^+/(e^++e^-)$ above 10~GeV up to 100~GeV~\cite{PAMELApositrons}, compatibly with previous less certain hints from HEAT~\cite{HEAT} and AMS-01~\cite{AMS01};
\item[$\cdot$] no excess in the $\bar p/p$ energy spectrum~\cite{PAMELApbar} compared with the predicted background, up to the maximal probed energy of about 100 GeV.
\end{itemize}
If interpreted in terms of Dark Matter annihilations, these results indicate~\cite{CKRS} either 
\begin{itemize}
\item[(i)] a DM particle $\chi$ of any mass (above about 100 GeV) that annihilates only into leptons ($\chi\bar\chi \to e^+e^-,\mu^+\mu^-,\tau^+\tau^-$), not producing therefore unseen antiprotons or 
\item[(ii)] a DM particle with a mass around or above a few TeV, that can annihilate into any channel (i.e. $\chi\bar\chi \to W^+W^-, ZZ, b\bar b, t\bar t$, light quark pairs and the leptonic channels above) possibly producing anti-proton fluxes at energies above those currently probed by PAMELA.
\end{itemize}
We will of course concentrate on the second possibility in this paper, as annihilations into leptonic channels do not produce any anti-nucleons and a fortiori no anti-deuterium. Instead, annihilations into gauge bosons or quarks copiously produce anti-proton and anti-neutron fluxes and therefore entail a possibly large production of anti-deuterons. We will consider Dark Matter masses $M_{\rm DM}$ equal and above 1 TeV as indicated in table \ref{tab:Msigma}, being aware that the smallest value (1 TeV) is kept mainly for reference purposes, as it is already excluded by the non-observation of anti-protons.\\
The PAMELA results also require a very large annihilation cross section, in order to match the large observed flux in positrons. Based on the analysis of~\cite{CKRS} we adopt the benchmark values indicated in table~\ref{tab:Msigma}.
\begin{table}[t]
\center
\begin{tabular}{c|c}
DM mass $M_{\rm DM}$ [TeV] & $\langle \sigma v \rangle\, [{\rm cm}^3/{\rm sec}]$\\[2mm]
\hline
\hline\\[-3mm]
{\em 1} & \hspace{-1mm} {\em 5} $\cdot$ {\em 10}$^{\mbox{{\em {\scriptsize $-$23}}}}$\\
5 	&  $3 \cdot 10^{-22}$\\
10	&  $7 \cdot 10^{-22}$\\
20	&  $2 \cdot 10^{-21}$\\
\hline\\[-3.3mm]
9.6 & eq.\,(\ref{eq:annihilationsigma})$\cdot$(\ref{eq:MDMboost})
\end{tabular}
\caption{\em Masses and annihilation cross sections of a heavy DM particle that fits the PAMELA positron excess (see~\cite{CKRS}), rather independently from the annihilation channel. The 1 TeV case is quite disfavored by the antiproton results and is kept mainly for reference purposes. The last line refers the Minimal Dark Matter model reviewed in sec. \ref{MDMmodel}. \label{tab:Msigma}}
\end{table}
These very large cross sections, much larger than the typical annihilation cross section required by DM thermal production in cosmology, $\approx 3\cdot 10^{-26}\, {\rm cm}^3/{\rm sec}$~\cite{reviews}, can be justified in specific models in terms of some enhancement mechanism which is effective today but not in the early universe (such as a resonance~\cite{CKRS,resonance} or Sommerfeld~\cite{Sommerfeld,MDMastro,NimaNeal} enhancement, the presence of an astrophysical boost factor due to DM substructures --unlikely\,\cite{Lavalle}--, or a combination of these). For the purposes of our model independent approach, they are taken as an input required by data.

\medskip

In addition to the PAMELA data, the balloon-borne experiments ATIC~\cite{ATIC-2} and PPB-BETS~\cite{Torii:2008xu} have recently reported results indicating the presence of an anomalous peak in the spectrum of cosmic ray $e^++e^-$ at about 500-800 GeV. The HESS telescope has also published data~\cite{HESSleptons} in the range of energy from 600 GeV up to a few TeV, showing a steepening of the spectrum which is compatible both with the ATIC peak (which cannot however be fully tested) and with a feature power law with index $-3.05 \pm 0.02$ and a cutoff at $\approx$2 TeV. If interpreted in terms of DM annihilations, the ATIC peak would clearly pin down the DM mass at about 1 TeV, and as a consequence select the possibility (i) above, making the generation of anti-deuteron fluxes impossible. However, the interpretations of the results from these experiments have to rely on MonteCarlo simulations that are tested only up to much smaller energies. Moreover, the agreement between the different balloon experiments is not optimal, e.g. the EC data~\cite{EC} fail to clearly indicate the presence of the peak. We work therefore under the assumption that the possibility (ii) above is the correct one. The upcoming data from the FERMI satellite~\cite{FERMIleptons} will be crucial to settle the issue.

If the signals in PAMELA are due to DM annihilations, one should also worry about the associated emission of high energy gamma rays from the galactic center (GC), the galactic ridge and the Sagittarius dwarf galaxy, and also of radio waves from the galactic center (produced by synchrotron radiation in the strong magnetic field by the electrons and positrons from DM annihilations). These can impose stringent constraints on the annihilation cross section. A complete, model independent calculation has been carried out in this respect in~\cite{BCST}. It is found that, for most annihilation channels, the cross sections required by PAMELA are excluded if the DM distribution in the galaxies follows a benchmark NFW profile (reviewed below). But choosing a smoother profile and/or assuming that a part of the cross section is due to an astrophysical boost factor that would not be present in dwarf galaxies and the Galactic Center due to tidal disruption re-allows the cross sections of table \ref{tab:Msigma}. How realistic these choices and assumptions are will be clarified by further observations, studies (especially of dwarf galaxies DM distributions) and simulations.

\medskip

In summary: the recent PAMELA data select a heavy, multi-TeV DM particle with a large annihilation cross section (see table \ref{tab:Msigma}) as a possible candidate for explaining the excess in positrons and the null result in antiproton fluxes (the tension with the ballon experiment data in the $e^++e^-$ spectra has to be settled by further data and we do not consider it in the following, while the bounds from $\gamma$ rays and radio signals can be avoided with assumptions on the DM distribution that will be tested in the future).

\subsection{Minimal Dark Matter}
\label{MDMmodel}
A particular model featuring DM particles of the kind favoured by the PAMELA data is the Minimal Dark Matter (MDM) model \cite{MDM,MDMastro,MDMindirect,MDMidm08,MDMreview}. It consists of a minimalistic approach to solve the dark matter problem, not inspired by more ambitious Beyond-the-Standard-Model ideas such as supersymmetry or extra dimensions. It is constructed by adding to the SM one new $n$-plet of the ${\rm SU}(2)_L$ gauge group and assigning quantum numbers such that a viable dark matter candidate is found: a stable neutral particle still allowed by current DM searches. To automatically insure stability, $n$-plets that possess decay modes into SM particles consistent with renormalizability are excluded. To keep it really minimalistic, scalar $n$-plets with quartic couplings to the SM higgs are rejected as well. The only interactions with SM particles are then of gauge type and the only new parameter is the tree level mass $M_{\rm MDM}$ of the new multiplet. This mass is fixed by the requirement that the thermal relic abundance equals the measured DM abundance. Direct searches for DM exclude all n-tuplets with $Y\neq0$ as they have interactions with the $Z$ boson that lead to an elastic scattering cross section with nucleons which is 2 - 3 orders of magnitude above present bounds from direct DM searches. All in all, a fermionic quintuplet with $Y=0$ and mass $M_{\rm MDM}=9.6\ {\rm TeV}$ is singled out as the most successful candidate.\\
MDM particles annihilate into $W^+W^-$ at tree level and into $ZZ$, $\gamma Z$ and $\gamma\gamma$ at one-loop. Due to the Sommerfeld enhancement, the cross sections for these annihilations are much higher than what is normally expected for WIMPs. One finds~\cite{MDMastro}:
\beq
\label{eq:annihilationsigma}
\langle \sigma v \rangle_{WW} = 1.1\cdot 10^{-23} {{\rm cm}^3 \over {\rm sec}},\quad
\langle \sigma v \rangle_{ZZ}= 3.2 \cdot 10^{-24} {{\rm cm}^3\over {\rm sec}},\quad
\langle \sigma v \rangle_{\gamma Z}= 6.5 \cdot 10^{-25} {{\rm cm}^3\over {\rm sec}}.
\eeq
Note that this is an example of a model in which the annihilation cross section is of the typical weak order of $3 \cdot 10^{-26}{{\rm cm}^3 / {\rm sec}}$ at thermal freeze out, and thus produces the right DM abundance $\Omega_{\rm DM} \simeq 6\cdot 10^{-27} \frac{{\rm cm}^3}{\rm sec}\, \langle \sigma v \rangle^{-1}$ (the so-called ``WIMP miracle''), but then acquires the larger values of eq.\eq{annihilationsigma} at later times. This is because the Sommerfeld enhancement is roughly inversely proportional to the particle velocity~\cite{MDMastro} and the particles in the early universe had much higher velocities ($\beta \sim 0.1$) than in the galactic halo today ($\beta \sim 10^{-3}$).\\
If a boost factor due to DM overdensities in the halo
\beq
B \simeq 50
\label{eq:MDMboost}
\eeq 
is assumed\footnote{We stick to these given values of $\sigma v$ and $B$ for definiteness in our analysis of MDM, but we note that such values can vary by about one order of magnitude --thus possibly reducing $B$ to about 5 and correspondingly increasing $\sigma v$-- within the 3 $\sigma$ range of $\Omega_{\rm DM}$ (see~\cite{MDMindirect,MDMreview}). A larger $\sigma v$ exacerbates the tension with gamma ray constraints from the galactic center and dwarf galaxies, but a smaller boost would be preferable for the agreement with numerical simulations.}, the predictions of the MDM model for an excess of positrons in cosmic rays due to DM annihilations are well matched by the PAMELA results (see addendum to \cite{MDMindirect} and \cite{MDMreview}). 

\section{Anti-deuteron production and propagation in the galaxy}
\label{prodandprop}

\subsection{Primary fluxes from DM annihilations}
\label{primary}

We consider the annihilation of a dark matter particle $\chi$ with its antiparticle $\overline{\chi}$ into pairs of bosons and quarks:
 \begin{equation}
  \chi\overline{\chi} \rightarrow W^+W^-, \gamma Z, ZZ, c\bar{c}, b\bar{b}, t\bar{t},\\
 \end{equation}
where \(\chi\overline{\chi}\rightarrow c\bar{c}\) stands representative for annihilations into light quarks. 
%In the special case of Minimal Dark Matter, the decay into $ZZ$ and \(\gamma Z\) is a one-loop process and the decay channels into quarks are highly suppressed as MDM cannot couple to Z (because Y=0 for all MDM candidates not already excluded by direct searches, see above) \cite{MDM}. 
Of course, other annihilation channels, such as \(\chi\overline{\chi}\rightarrow\gamma\gamma\) are also conceivable but would not produce any \(\bar{p}\) and \(\bar{n}\) and thus no \(\bar{d}\).

\medskip

Extending the SM to incorporate DM particles, we simulate these processes using the MadGraph 4.4 Monte Carlo event generator \cite{MadGraph} for the masses of the DM particles $\chi$ indicated in table \ref{tab:Msigma}. We then hand over the resulting bosons and quarks to PYTHIA 8.1 \cite{PYTHIA} for showering and hadronization, explicitely requesting the \(\bar{n}\) not to decay, as they are needed for the formation of \(\bar{d}\). The produced $\bar{p}$-fluxes as a function of the kinetic energy $T$ agree very well with the fluxes presented in \cite{Donato:2003xg}\footnote{Our fluxes are of course a factor of 2 lower as we keep $\bar n$ from decaying. We note that the fluxes in their fig.3 most probably refer to $dN/dx$ and not $dN/dT$ as indicated.}. The fluxes of $\bar{n}$-fluxes are essentially equal to the $\bar{p}$ ones, as expected from isospin invariance.

\subsection{Anti-deuteron formation by coalescence}
\label{coalescence}

Anti-deuterons are then produced via the fusion process of a $\bar p$-$\bar n$ pair. This is described by the very simple coalescence model~\cite{coalescence}.
In this model the so-called ``coalescence function'' $\mathscr{C} (\vec{k}_{\bar{p}},\vec{k}_{\bar{n}})$ describes the probability for a \(\bar{p}\)-\(\bar{n}\) pair to yield by fusion an antideuteron. It actually depends only on the difference of the momenta \(\vec{k}_{\bar{p}}-\vec{k}_{\bar{n}}=2\vec{\Delta}\) and is strongly peaked around \(\vec{\Delta}=0\):
\begin{equation}
 \vec{k}_{\bar{p}}\approx\vec{k}_{\bar{n}}\approx\frac{\vec{k}_{\bar{d}}}{2}.
\end{equation}
The \(\bar{d}\)-density in momentum space is thus written as the \(\bar{p}\)-density (in momentum space) times the probability to find an \(\bar{n}\) within a sphere of radius p$_{0}$ around \(\vec{k}_{\bar{p}}\):
\begin{equation}
 \underbrace{\gamma_{\bar{d}}\frac{d^{3}N_{\bar{d}}}{d\vec{k}_{\bar{d}}^3}}_{\substack{\bar{d}\mbox{\footnotesize -density in}\\ \mbox{\footnotesize momentum space}}}=\underbrace{\frac{4\pi}{3}p_{0}^{3}\gamma_{\bar{n}}\frac{d^{3}N_{\bar{n}}}{d\vec{k}_{\bar{n}}^{3}}}_{\substack{\mbox{\footnotesize probability to find }\\\bar{n}\mbox{\footnotesize \ within a sphere}\\ \mbox{\footnotesize of radius }p_{0} \mbox{\footnotesize \, around }\vec{k}_{\bar{p}}\\ \mbox{\footnotesize in momentum space}}}\cdot\underbrace{\gamma_{\bar{p}}\frac{d^{3}N_{\bar{p}}}{d\vec{k}^{3}_{\bar{p}}}}_{\substack{\bar{p}\mbox{\footnotesize  -density in}\\ \mbox{\footnotesize  momentum} \mbox{\footnotesize\, space}}}
\end{equation}
where $\gamma$ is the Lorentz factor and the ``coalescence momentum'' $p_{0}$ is a free parameter constrained by data on hadronic production~\cite{Aktas:2004pq,Chekanov:2007mv,Schael:2006fd}. 
% (it is of course assumed that the same value holds for final products from DM annihilations, and there is no reason to believe in the contrary).
The central value preferred in~\cite{Duperray:2005si, Donato2008} is $p_0 = 79$ MeV, while a window of  66 - 105 MeV is given in~\cite{Donato2008}. It is energy independent for a wide range of energies~\cite{Duperray:2005si,Aktas:2004pq}.\\
In brief, one assumes that the antinucleons merge whenever they are within a momentum \(p_{0}\) of each other  \cite{Kapusta:1980zz, Chardonnet:1997dv, Donato2008}. After some algebra, one gets the following formula for the \(\bar{d}\)-spectrum at the site of production, where $T$ denotes the kinetic energy:
\begin{equation}
 \frac{dN_{\bar{d}}}{dT_{\bar{d}}}=\frac{4}{3}p_{0}^{3}\frac{m_{\bar{d}}}{m_{\bar{n}}m_{\bar{p}}}\frac{1}{\sqrt{T_{\bar{d}}^{2}+2m_{\bar{d}}T_{\bar{d}}}}\left(\frac{dN_{\bar{p}}}{dT_{\bar{p}}}\right)_{T_{\bar{p}}=\frac{T_{\bar{d}}}{2}}\left(\frac{dN_{\bar{n}}}{dT_{\bar{n}}}\right)_{T_{\bar{n}}=\frac{T_{\bar{d}}}{2}}
\end{equation}
Notice that we apply the coalescence prescription to the fluxes of $\bar p$ and $\bar n$ as produced by the MonteCarlo code. A more precise computation, that we do not perform, would require to coalesce the $\bar p$ and $\bar n$ pair {\it event per event} within the MonteCarlo itself.

\begin{figure}[t]
\begin{center}
\includegraphics[width=0.475\textwidth]{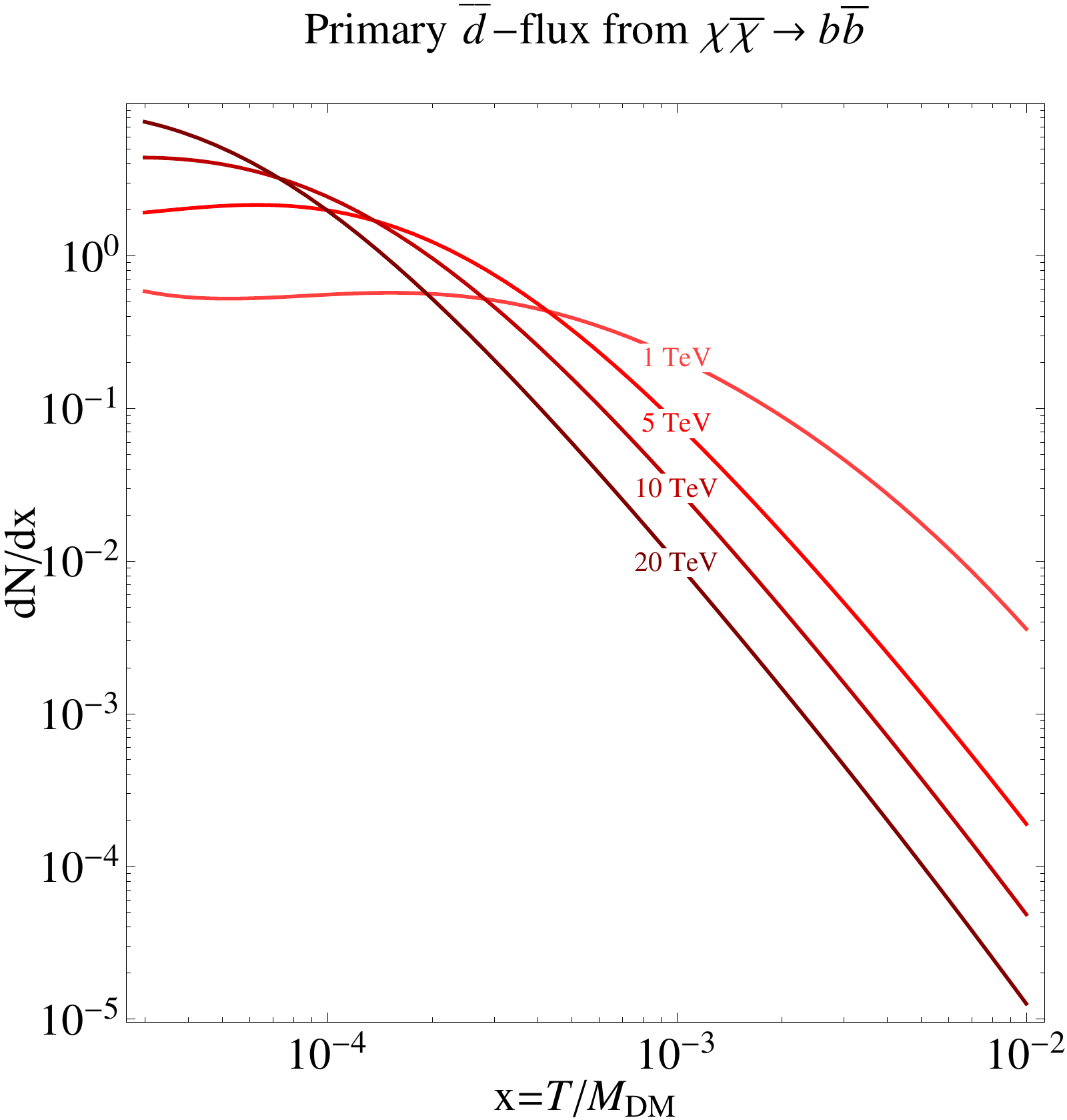}
\includegraphics[width=0.475\textwidth]{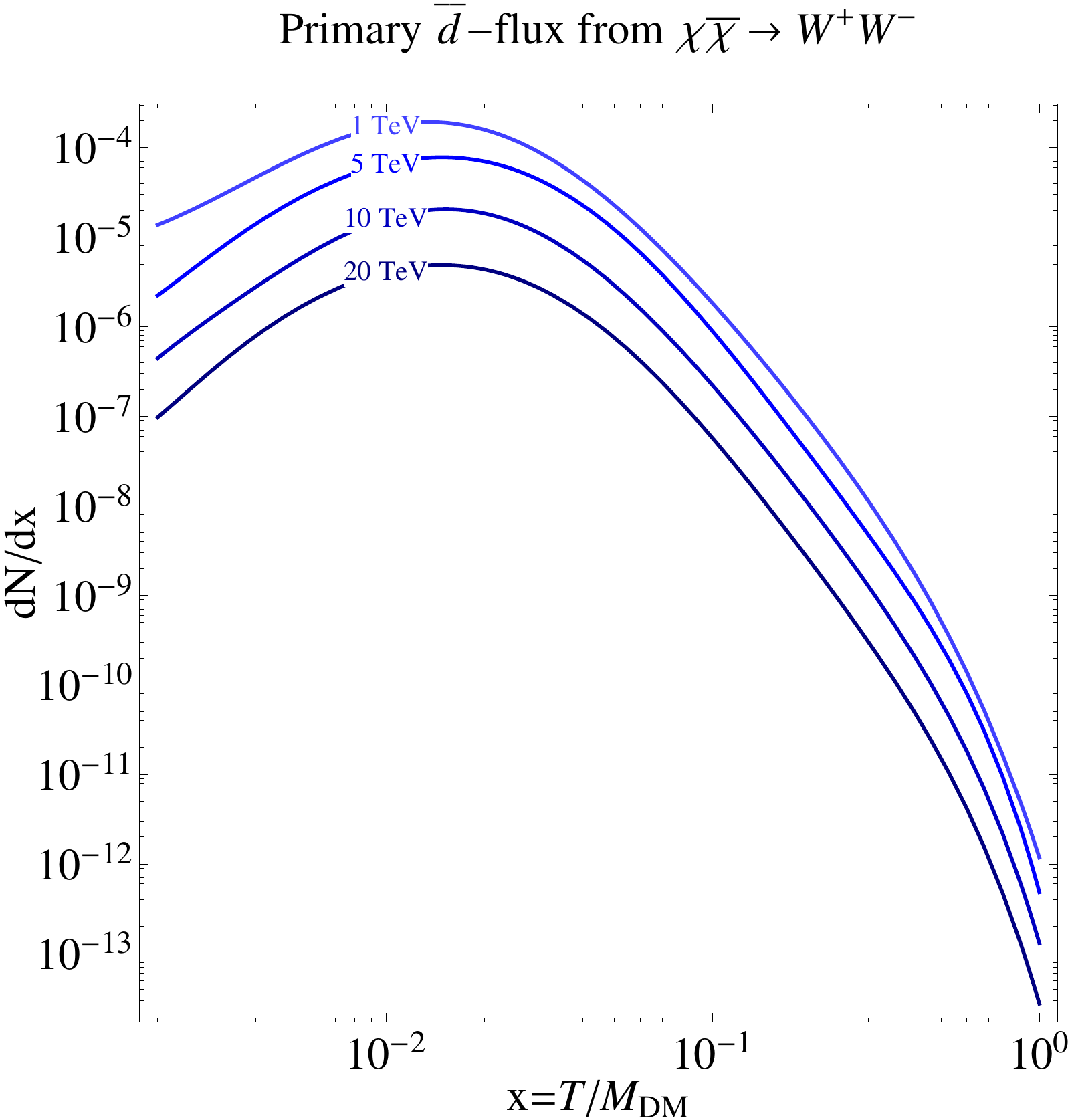}
\caption{\em\label{fig:primary} Spectra of antideuterons per annihilation from DM annihilations in the indicated primary channels, for the masses of table \ref{tab:Msigma}.}
\end{center}
\end{figure}

Fig.\fig{primary} shows our results for the primary antideuteron fluxes. In fact, all quark channels lead to fluxes that are quite similar among each other, as do respectively the gauge boson channels. We therefore select here and in the following as exemplary the cases of the $b\bar{b}$ and the $W^+W^-$ channels.
One sees that the gauge boson channels lead to fluxes that are less abundant and concentrated at high energies. This   allows to foresee a less promising signal (after propagation) with respect to the one from primary quarks.

\subsection{Propagation}
\label{propagation}
Once the antideuterons have been produced, in each given point of the DM halo, they propagate in the galactic region and eventually reach the Earth. In this section we review the processes to which they undergo during their trip and the formalism that allows to compute the resulting final spectra.

Assuming steady state, the diffusion-loss equation describing the change in the differential number density \(\mathcal{N}(r,z,T)\equiv\frac{dN}{dT}\) of cosmic ray particles when travelling through the galaxy in cylindrical coordinates $r,z$ is \cite{Maurin:2001sj, Donato:2001ms,Maurin:2002ua, Barrau:2001ev, Sendouda:2004hz}
\begin{equation}
\begin{split}
\vec{\nabla}[\underbrace{K(T)\vec{\nabla}\mathcal{N}(r,z,T)}_{\mbox{\footnotesize{(spatial) diffusion}}}-\underbrace{\vec{V}_{c}(r,z)\mathcal{N}(r,z,T)}_{\mbox{\footnotesize{convection}}}\ ]\ -\ \underbrace{
\Gamma(T)\mathcal{N}(r,z,T)}_{\mbox{\footnotesize{annihilation}}}\ +\ \underbrace{\mathscr{Q}(r,z,T)}_{\mbox{\footnotesize{source}}}\\
+\underbrace{\frac{\vec{\nabla}\cdot\vec{V}_{c}(r,z)}{3}\frac{\partial}{\partial T}\left[\frac{p^{2}}{T}\mathcal{N}(r,z,T)\right]}_{\mbox{\footnotesize{adiabatic loss due to galactic wind}}}-\underbrace{\frac{\partial}{\partial T}\left[\left(b_{\rm ion}(T)+b_{\rm Coul}(T)\right)\mathcal{N}(r,z,T)\right]}_{\mbox{\footnotesize{ionization and Coulomb losses}}}\\-\ \underbrace{\frac{\partial}{\partial T}\left[\frac{1+\beta^{2}}{T}K_{pp}(T)\mathcal{N}(r,z,T)\right]\ +\ \frac{\partial}{\partial T}\left[\beta^{2}K_{pp}(T)\frac{\partial}{\partial T}\mathcal{N}(r,z,T)\right]}_{\mbox{\footnotesize{first \& second order reacceleration}}}&=0
\end{split}
\label{eq:DiffN}
\end{equation}
The last four terms are energy redistribution terms. They can be neglected~\cite{fastformulae} for the fluxes of primary anti-deuterons. The basic physics reason is that all these processes happen in the thin disk of the galaxy, where the anti-deuterons spend little time (they rarely cross the galactic plane). We keep however the annihilation terms that have the more drastic effect of removing a particle from the flux\footnote{This result applies to $\bar p$ as well. Notice that positrons, also used for the indirect search for Dark Matter, have energy loss terms corresponding to Inverse Compton and bremsstrahlung: these effects are not relevant for $\bar d$ (and $\bar p$) since $m_{\bar d} \gg m_{e^+}$.}.
The same approximation is used in~\cite{Donato:2003xg}, where the procedure to re-include those terms is also briefly discussed.
%\subsubsection{The two-zone disk-halo model}

To be able to analytically solve the diffusion equation\eq{DiffN} one has to employ  a simplified description of the spatial distribution of matter in the Galaxy. The so-called two-zone disk-halo model is adopted, where the Galaxy is described as a thin gaseous disk (of radius $R=20$ kpc and half-height $h=100$ pc) embedded in a thick, cylindrical diffusive halo. The half-height $L$ of the halo is a free parameter of the model, constrained from the study of Boron/Carbon (B/C) CR ratios \cite{Maurin:2001sj,Maurin:2002hw}.
%\subsubsection{Diffusion and convection}

Spatial diffusion arises because charged particles interact with the galactic magnetic field inhomogenities.  Diffusion is an energy-dependent process as higher energy particles probe larger spatial scales than lower energy ones. The energy-dependent diffusion coefficient is given by \cite{Maurin:2001sj,Sendouda:2004hz}
\begin{equation}
K(E)=K_{0}\beta \,\mathcal{R}^{\delta}
\label{diffusion_coefficient}
\end{equation}
where \(\beta\) as usual denotes the velocity in units of $c$ and the rigidity, \(\mathcal{R}=\frac{|\vec{p}|c}{Ze}\) for a particle of atomic number $Z$. \(K_{0}\) and \(\delta\) are again constrained by B/C data. \\
The presence of a galactic wind \(\vec{V}_{c}\) of cosmic rays pointing outwards from the galactic plane is a well established fact. This leads to convection and an energy loss due to the adiabatic expansion of the plasma (neglected here), as described in the diffusion equation\eq{DiffN}. The strength of this wind, \(V_{c}\), is again a parameter to be determined by fitting to B/C-data.

The transport parameters \(L, K_{0}, \delta, V_{c}\) providing the maximal, median and minimum primary antideuteron flux compatible with B/C analysis are reported in table \ref{TransportParameters}. We will later study the effect on the propagated fluxes of choosing one or the other of the different sets.

\medskip 

\begin{figure}[t]
\begin{center}
\includegraphics[width=0.6\textwidth]{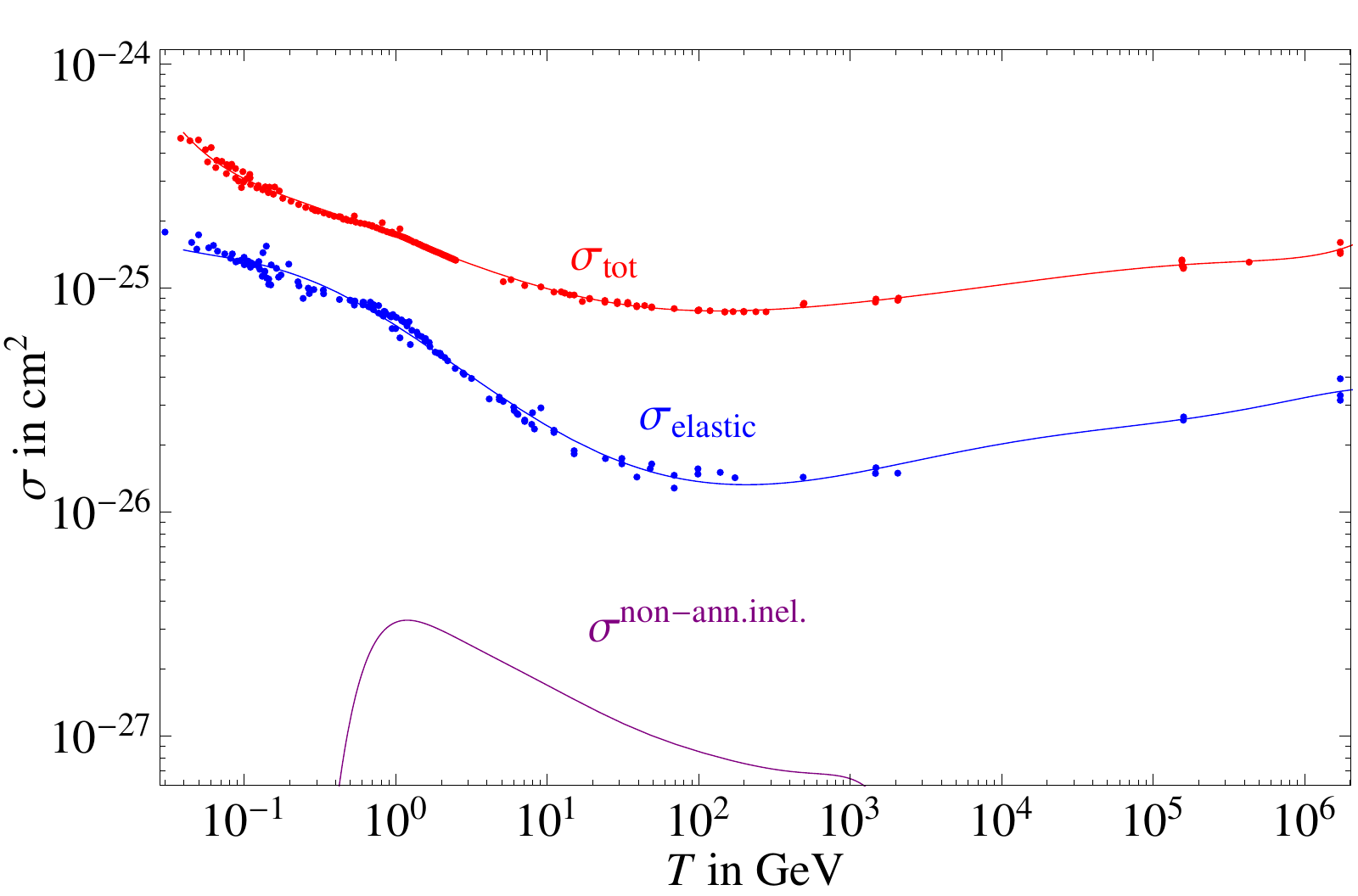}
\caption{\em\label{fig:nuclear} Cross sections for the different processes of interaction of light nuclei on the interstellar gas, and the experimental datasets on which they are based.}
\end{center}
\end{figure}

%\subsubsection{Annihilation on ISM}
The third term in eq.\eq{DiffN} describes the annihilation of the anti-deuterons with H and He in the interstellar medium (ISM) in the thin galactic disk. \(\Gamma(T)\) is therefore defined as:
\begin{equation}
 \Gamma(T)=2h\delta(z)\Gamma^{\rm ann}(T),
 \label{ann}
\end{equation}
where
\begin{equation}
 \Gamma^{\rm ann}(T)=\left(n_{\rm H}+4^{\frac{2}{3}}n_{\rm He}\right)v\sigma^{\rm ann}_{\rm inel}(T),
\end{equation}
with $n_H \approx 1/{\rm cm}^3$ the hydrogen density and $n_{\rm He}\approx 0.07\, n_{\rm H}$ the Helium density. 
In order to calculate
\begin{equation}
 \sigma^{\rm ann}_{\rm inel}=\sigma_{\rm inel}-\sigma^{\rm non-ann}_{\rm inel}=\sigma_{\rm tot}-\sigma_{\rm el}-\sigma^{\rm non-ann}_{\rm inel},
\end{equation}
where \(\sigma^{\rm non-ann}_{\rm inel}\) corresponds to anti-deuterons interacting inelastically with the ISM but surviving the collision, loosing only a fraction of their energy, one needs the total, elastic and non-annihilating inelastic cross sections for the process $\bar{d}+H$ \cite{Donato2008}. 
%\xxx{$\sigma^{non-ann}_{inel}$ neglected here???? I think we don't need to neglect that we can extract the data up to $10^{3}$ GeV from Donato et al. and there is no need for higher energies.}
For what concerns the total cross-section, no experimental data exist, but data for the charge conjugate reaction \(d\bar{p}\) can be obtained from the \href{http://pdg.lbl.gov/2008/hadronic-xsections/hadron.html}{PDG} \cite{Amsler:2008zzb}. We assume therefore 
\begin{equation}
 \sigma^{\bar{d}p}_{\rm tot}\approx\sigma^{d\bar{p}}_{\rm tot}.
\end{equation}
As for the elastic cross section, no data are available even for the charge conjugate reaction. We therefore make the assumption
\begin{equation}
 \sigma_{\rm el}^{p\bar{d}}\approx2\sigma_{\rm el}^{p\bar{p}},
\end{equation}
and data on \(\sigma_{el}^{p\bar{p}}\) can again be obtained from the \href{http://pdg.lbl.gov/2008/hadronic-xsections/hadron.html}{PDG}. At energies $\gtrsim10^{3}$ GeV, we make this assumption for the total cross section as well. 
As it would be complicated to calculate the cross section \(\sigma_{\rm inel}^{\rm non-ann}=\sigma(\bar{d}p\rightarrow\bar{d}X)\) due to the complex interplay of the underlying NN cross sections and selection rules, it is obtained from experimental values (of the symmetric system \(\bar{p}d\rightarrow Xd\) as there are - again - no experimental values for the reaction \(\bar{d}p\rightarrow\bar{d}X\)). \(\sigma(\bar{p}d\rightarrow Xd)\) is found by summing up the \(\bar{p}d\rightarrow(n\pi)\bar{p}d\) cross sections experimentally available. Channels not known experimentally are assumed to give only negligibly small contributions \cite{Duperray:2005si}. Figure \ref{fig:nuclear} illustrates the different ingredients that we have just described.
%\xxx{If we put all this detailed discussion, we should put at least a plot of the sigmas. Do you have that?}
%\xxx{I write nothing on energy gain and losses as we neglect that anyway, o.k.?}

%\subsubsection{Sources}

The source term contains a primary term (the signal from DM), a secondary term (the astrophysical background) and a tertiary term
\begin{equation}
 \mathscr{Q}(r,z,T)=Q^{\rm prim}(r,z,T)+Q^{\rm sec}(r,z,T)+Q^{\rm tert}(r,z,T).
\end{equation}
The tertiary source term merely describes the fact that some antideuterons move to higher energy bins by interaction with the ISM, others to lower energy bins. As already stated above we neglect this energy redistribution. 
%We solve the diffusion equation for the DM primary source only (we take the background ).
%Annihilations of DM particles occur throughout the disk and halo with the same probability. 
We will take the propagated background fluxes from the literature (see below), so we only solve the equation for the primary source
\begin{equation}
 Q^{\rm prim}(r,z,T)=\eta \left(\frac{\rho_{\rm DM}(r,z)}{M_{\rm DM}}\right)^{2}\sum_{k} \langle\sigma v\rangle_{k}\frac{dN^{k}_{\bar{d}}}{dT}.
\end{equation}
The sum is over annihilation channels with \(\bar{p}, \bar{n}\) in the final state and annihilation cross section of the DM particles $\langle \sigma v\rangle_{k}$ and $dN^{k}_{\bar{d}}/dT$ are the source spectra computed in sec.~\ref{primary}. The coefficient \(\eta\) depends on the DM particle being or not self-conjugate: for example, for fermions it is 1/2 for Majorana and 1/4 for Dirac particles. For definiteness (and also for consistency with the specific case of  Minimal Dark Matter) we choose to focus on  the Majorana case. The Dirac case just amounts to rescaling our results by a factor 2.\\ 
For the distribution of DM in the galaxy, \(\rho_{\rm DM}(r,z)\), we consider, as customary, three different models determined by numerical simulations. The \textit{Navarro-Frenck-White} profile~\cite{Navarro:1995iw}
\begin{equation}
 \rho_{\rm NFW}(r)=\rho_{s}\frac{r_{s}}{r}\left(1+\frac{r}{r_{s}}\right)^{-2},
\end{equation}
the \textit{Einasto} profile~\cite{Graham:2005xx, Navarro:2008kc}
\begin{equation}
 \rho_{\rm Einasto}(r)=\rho_{s}\exp\left[-\frac{2}{\alpha}\left(\left(\frac{r}{r_{s}}\right)^{\alpha}-1\right)\right],\ \ \alpha=0.17,
\end{equation}
and the \textit{isothermal} profile~\cite{Bahcall:1980fb}
\begin{equation}
 \rho_{\rm iso}(r)=\frac{\rho_{s}}{1+\left(\frac{r}{r_{s}}\right)^{2}}.
\end{equation}
The values for the parameters \(r_{s}\) and \(\rho_{s}\) of the three models are given in table \ref{DMprofiles}. While they  sensibly differ at the Galactic Center (from the cored isothermal profile to the more cuspy NFW), they do not at the proximity of the Earth. We will later study the effect on the propagated fluxes of changing the choice of profile.

\begin{table}
 \begin{minipage}[t]{0.45\textwidth}
 \footnotesize{
 \renewcommand{\arraystretch}{1.5}
 \centering
  \begin{tabular}{c|c|c|c|c}
   case&L [kpc]&$\delta$&$K_{0}$ $\left[\frac{kpc^{2}}{Myr}\right]$&$V_{c} \left[\frac{km}{s}\right]$\\
   \hline
   max&15&0.46&0.0765&5\\
   med&4&0.7&0.0112&12\\
   min&1&0.85&0.0016&13.5
  \end{tabular}
 \caption{Transport parameters providing the maximal, median and minimal primary antideuteron flux compatible with B/C data~\cite{Donato2008}}.
 \label{TransportParameters}}
\end{minipage}
\qquad
\begin{minipage}[t]{0.45\textwidth}
\footnotesize{
 \renewcommand{\arraystretch}{1.5}
\centering
 \begin{tabular}{c|cc}
  DM halo model & \(r_{s}\) in kpc & \(\rho_{s}\) in \(\frac{GeV}{cm^{3}}\)\\
  \hline
  NFW \cite{Navarro:1995iw} & 20 & 0.26\\
  Einasto \cite{Graham:2005xx, Navarro:2008kc} & 20 & 0.06\\
  Isothermal \cite{Bahcall:1980fb} & 5 & 1.16
 \end{tabular}
\caption{Parameters of the density profiles for the Milky Way DM halo.}
\label{DMprofiles}}
\end{minipage}

\end{table}

%\subsubsection{Analytical solution without energy redistribution}

\bigskip

Neglecting energy gain and loss terms, using cylindrical coordinates and Bessel-expanding \(\mathcal{N}(r,z,T)\) and \(Q^{\rm prim}(r,z,T)\)
\begin{equation}
 \mathcal{N}(r,z,T)=\sum_{i=1}^{\infty}N_{i}(z,T)J_{0}\left(\xi_{i}\frac{r}{R}\right)
\end{equation}
where \(\xi_{i}\) is the ith zero of \(J_{0}(x)\), the equation to solve in the halo is \cite{Maurin:2001sj, Donato:2001ms,Maurin:2002ua, Barrau:2001ev, Sendouda:2004hz}
\begin{equation}
 \left(\frac{\partial^{2}}{\partial z^{2}}-\frac{V_{c}}{K(T)}\frac{\partial}{\partial z}-\frac{\xi_{i}^{2}}{R^{2}}\right)N_{i}(z,T)=-\frac{Q_{i}^{\rm prim}(z,T)}{K(T)}.
\end{equation}
Imposing the boundary condition \(N_{i}(z=L)=0\) and ensuring continuity with the solution in the disk, where the equation to solve is
\begin{equation}
\left(\frac{\partial^{2}}{\partial z^{2}}-\frac{V_{c}}{K(T)}\frac{\partial}{\partial z}-\frac{\xi_{i}^{2}}{R^{2}}\right)N_{i}(z,T)=\frac{2h\Gamma^{ann}(T)}{K(T)}\delta(z)N_{i}(0,T)
\end{equation}
(there are no sources in the thin disk, but there is annihilation on the ISM, described by \(\Gamma^{\rm ann}(T)\)), 
one gets the solution\footnote{With
	\begin{equation}
	 y_{i}(z)=2\int_{0}^{z}\exp\left(\frac{V_{c}}{2K}(z-z')\right)\sinh\left(\frac{S_{i}}{2}(z-z')\right)Q^{\rm prim}_{i}(z')dz'
	\end{equation}
	\begin{minipage}{0.5\linewidth}
	\begin{equation}
	 A_{i}=2h\Gamma^{\rm ann}+V_{c}+KS_{i}\cdot \coth\left(\frac{S_{i}L}{2}\right)
	\end{equation}
	\end{minipage}
	\begin{minipage}{0.495\linewidth}
	\begin{equation}
	 S_{i}=\sqrt{\frac{V_{c}^{2}}{K^{2}}+4\frac{\xi^{2}_{i}}{R^{2}}}
	\end{equation}
	\end{minipage}
}
\begin{equation}
 \begin{split}
 N_{i}(z,T)=&\exp\left(\frac{V_{c}}{2K(T)}(z-L)\right)\frac{y_{i}(L)}{A_{i}\sinh\left(\frac{1}{2}S_{i}L\right)}\\ &\times\left[\frac{2h\Gamma^{\rm ann}+V_{c}}{S_{i}K(T)}\sinh\left(\frac{S_{i}z}{2}\right)+\cosh\left(\frac{S_{i}z}{2}\right)\right]-\frac{y_{i}(z)}{S_{i}K(T)}
 \end{split}
 \label{PropFlux}
\end{equation}
Finally, for the flux \(\Phi(r,z,T)=\frac{v}{4\pi}\mathcal{N}(r,z,T)\) near the Earth one therefore gets
\begin{equation}
\begin{split}
 \Phi(r=r_{\rm Sun}, z=0, T)&=\frac{B}{4\pi}c\sqrt{1-\frac{m_{\bar{d}}^{2}}{(T+m_{\bar{d}})^{2}}}\\ &\times\sum_{i=1}^{\infty}\exp\left(-\frac{V_{c}}{2K(T)}L\right)\frac{y_{i}(L)}{A_{i}\sinh\left(\frac{1}{2}S_{i}L\right)}J_{o}\left(\xi_{i}\frac{r_{\rm Sun}}{R}\right)
\end{split}
\end{equation}
where $B$ is the boost factor.

%\subsubsection{Solar modulation}

The last step needed is to include the average solar modulation effect: the solar wind decreases the kinetic energy $T$ of charged cosmic rays such that the flux \(\Phi_{\rm Earth}\) (denoted as TOA, for Top Of the Atmosphere) of anti-deuterons that reach the Earth with kinetic energy \(T_{\rm Earth}\) and momentum \(p_{\rm Earth}\) is approximately related to their flux in the interstellar medium as \cite{SolarMod}
\begin{equation}
 \Phi_{\rm Earth}=\frac{p_{\rm Earth}^{2}}{p^{2}}\Phi,\;\;\;\;\;\;T=T_{\rm Earth}+\left|Ze\right|\phi_{\rm Fisk},\;\;\;\;\;\;p^{2}=2m_{d}T+T^{2}.
\end{equation}
In this effective formalism the so-called Fisk potential \(\phi_{\rm Fisk}\) parameterizes the kinetic energy loss. We assume a value of \(\phi_{\rm Fisk}=0.5\) GV, characteristic of a minimum of the solar cyclic activity, corresponding to the period in which most of the observations were done in the second half of the 90's and are being done now.

\begin{figure}[t]
\begin{center}
\includegraphics[width=0.475\textwidth]{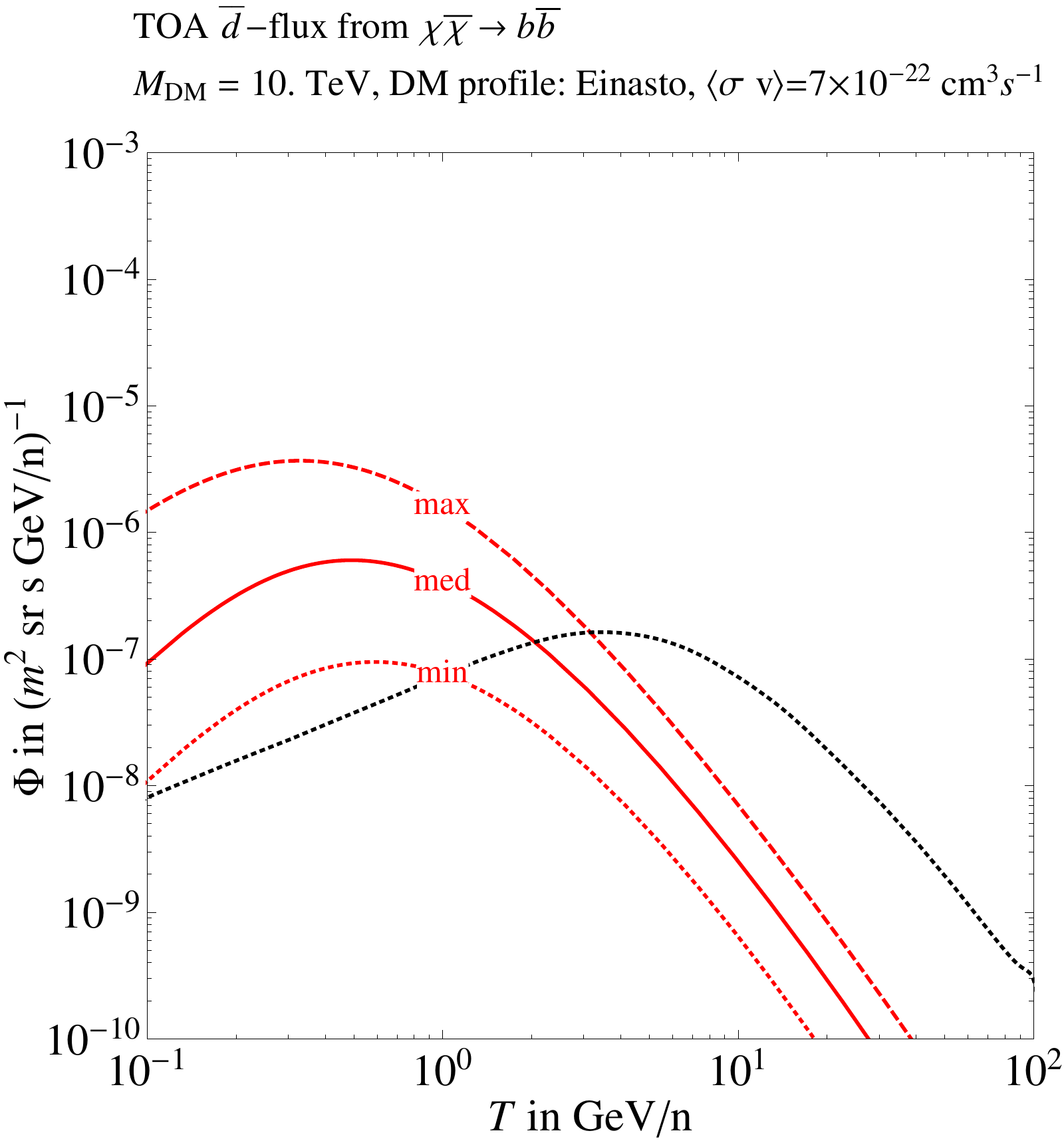}
\includegraphics[width=0.475\textwidth]{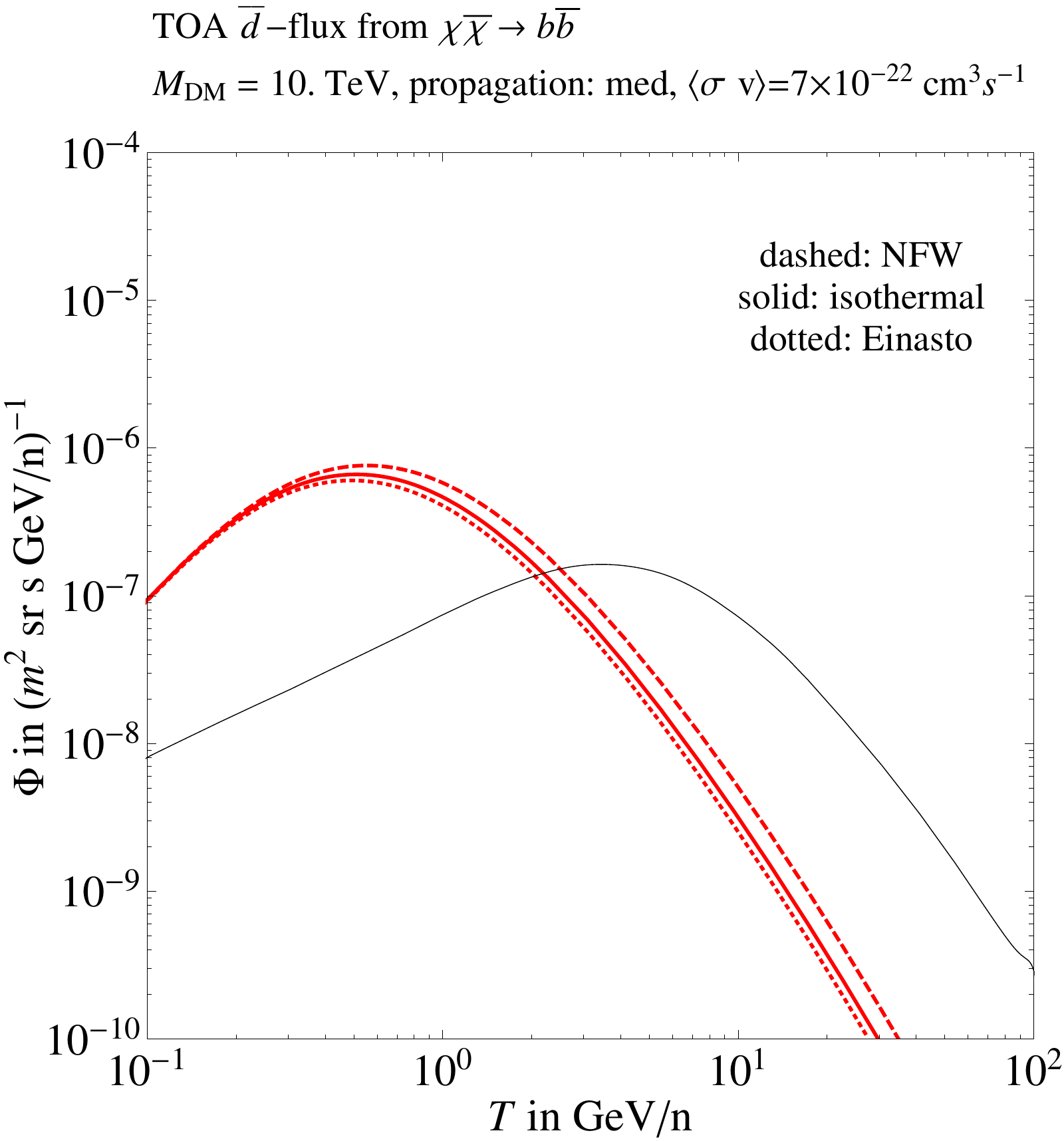}
\caption{\em\label{fig:prop} {\em Left panel}: Propagated (top-of-the-atmosphere (TOA)) $\bar{d}$ flux, including solar modulation, from a DM with mass 10 TeV annihilating into $b\bar{b}$, fixing the DM halo profile at  {\em Einasto} and changing the propagation parameters ({\em min, med, max}). {\em Right panel:} Fixing the propagation parameters at {\em med} and changing the halo profile ({\em NFW, Isothermal, Einasto}).}
\end{center}
\end{figure}

\begin{figure}[t]
\begin{center}
\includegraphics[width=0.475\textwidth]{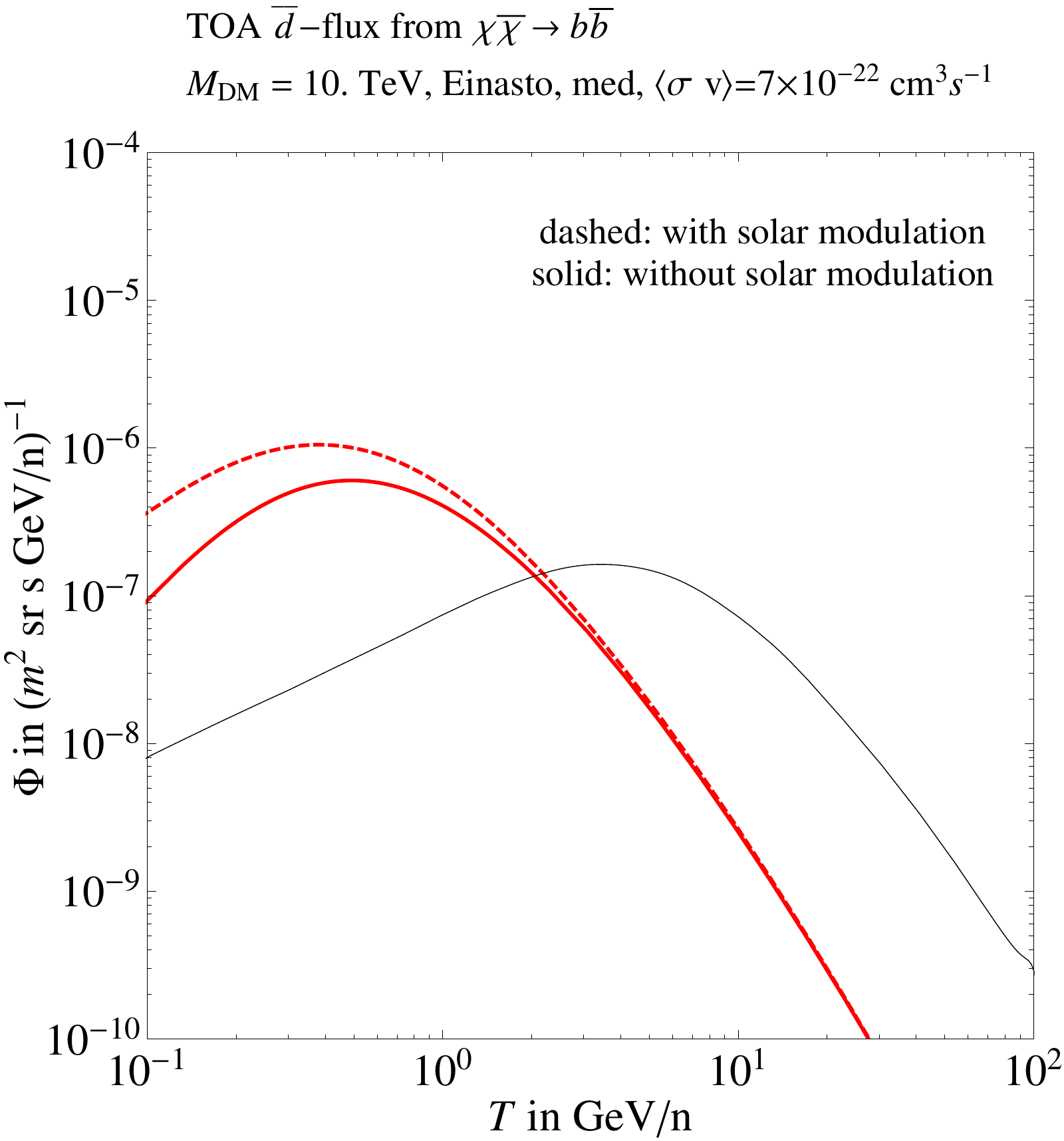}
\includegraphics[width=0.475\textwidth]{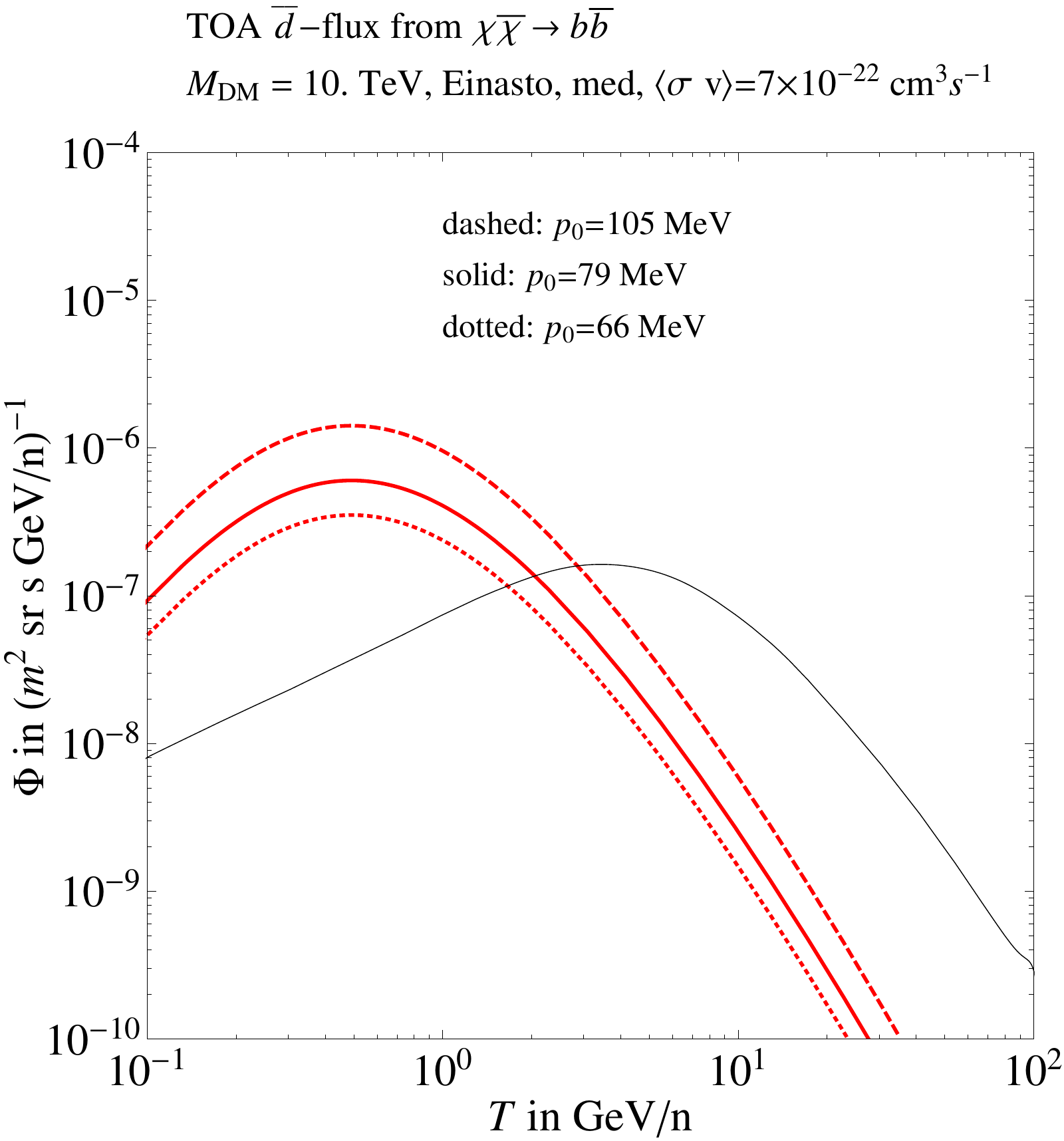}
\caption{\em\label{fig:prop2} {\em Left panel}: Propagated $\bar{d}$ flux from a DM with mass 10 TeV annihilating into $b\bar{b}$, fixing an {\rm Einasto} DM profile and {\em med} propagation parameters, with and without the solar modulation effect. {\em Right panel:} The same flux (including solar modulation) for different choices of the coalescence momentum $p_0$.}
\end{center}
\end{figure}

\section{Results and discussion}
\label{results}

\begin{figure}[t]
\begin{center}
\includegraphics[width=0.8\textwidth]{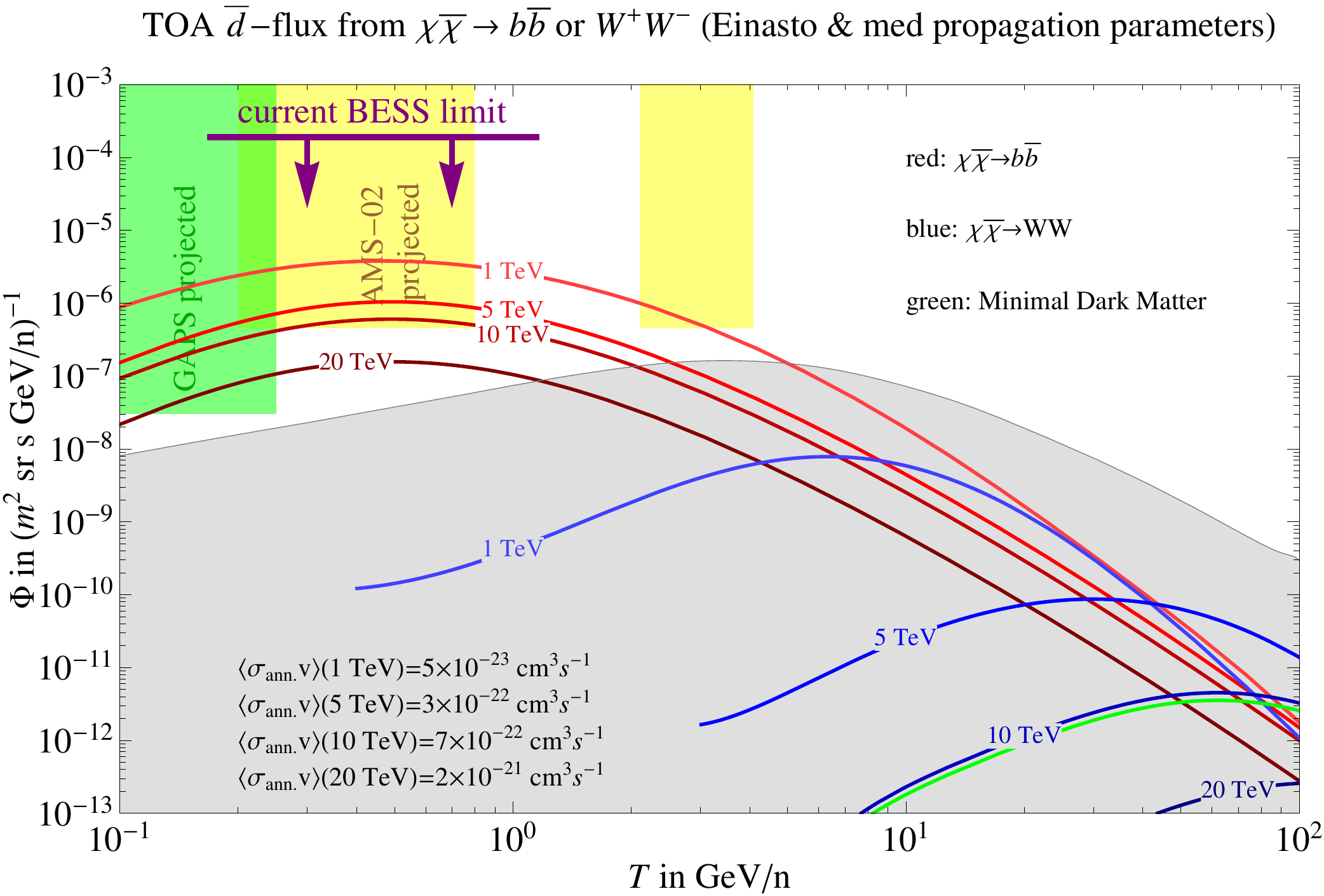}
\caption{\em\label{fig:final} The final fluxes of anti-deuterons for the different heavy DM candidates discussed in the paper and for Minimal Dark Matter compared to the background (gray shaded area) and to expected experiment sensitivities. Here we have assumed an {\em Einasto} DM profile, {\em med} propagation parameters and included the solar modulation effect.}
\end{center}
\end{figure}

In fig.s ~\fig{prop} -- \fig{final} we show our results. Fig.s \fig{prop} and \fig{prop2} illustrate the impact of the different astrophysical choices for one sample DM of $M_{\rm DM} = 10$ TeV annihilating into $b\bar{b}$, while fig.\fig{final} summarizes our main results for all the considered DM candidates and also reports the indication of the current experimental results and future sensitivities.

\bigskip

The background estimation reported in all our results is taken from the most recent calculation of~\cite{Donato2008}, fig.1, which agrees and slightly updates the detailed analysis in~\cite{Duperray:2005si}. Background $\bar d$ are produced by the spallation of high-energy cosmic ray (CR) $p$, He and $\bar p$ on the interstellar (IS) gas made of H and He in the galactic disk: the dominant source is CR $p$ on IS H for energies above 1 GeV, CR $\bar p$ on IS H and He for lower energies. Tertiary antideuteron production is important in this case. Notice the characteristic shape that decreases at $E < {\rm few \ GeV}$, opening an interesting window for the emergence of possible DM signals, as anticipated in the Introduction. For definiteness, the propagation scheme of the secondary $\bar d$ is fixed at {\em med} and the solar modulation at solar minimum (although, in principle, one should adapt the propagation parameters to the ones considered for the primaries). The uncertainty on such background is quoted at about a factor of 2 in~\cite{Duperray:2005si}.

\medskip

As it can be inferred from fig.\fig{prop}a, the choice of propagation parameters has quite an important impact on the predicted fluxes: the differences can amount to almost 2 orders of magnitude at the lowest energies. This is consistent with the analogous results in the case of anti-proton fluxes (see e.g.~\cite{MDMindirect}).\\ 
The choice of halo profile, instead, has a very limited effect, as shown in fig.\fig{prop}b. This is understood in terms of the fact that essentially no antideuterons come from as far as the region of the Galactic Center where the profiles differ significantly. This result is also in analogy with the antiproton case.

The solar modulation effect, shown in fig.\fig{prop2}a, has a certain effect in the lowest energy portion of the spectrum (it amounts to about a factor of 2 below 1 GeV), while the choice of the coalescence momentum within the intervall quoted in sec.~\ref{coalescence} has of course a simple overall normalization effect that amounts to about a factor of 4.

All in all, it is evident that the uncertainties ascribed to astrophysics and nuclear physics are very important, spanning a few orders of magnitude in the normalization of the fluxes (the shapes are instead more robust). Future improvements on B/C CR measurements might allow to restrict the ranges for the propagation parameters, a dominant source of uncertainty for the predictions.

\bigskip

Searches for anti-deuteron fluxes have been performed by BESS~\cite{BESSlimit}, a balloon-borne magnetic spectrometer experiment that collected data between 1997 and 2000. The null result imposes an upper limit on the flux in the explored energy range at 
\beq
\phi^{\rm BESS}_{\bar d} < 1.9 \cdot 10^{-4} \ ({\rm m}^2\  {\rm sec}\  {\rm sr}\  {\rm GeV}/{\rm n})^{-1}, \qquad  0.17\ {\rm GeV/n} < T_{\bar d} < 1.15 \ {\rm GeV/n} \qquad (95\% \, {\rm C.L.}).
\eeq
Future experimental capabilities rely on the performances of the AMS-02 cosmic ray experiment, to be installed on the International Space Station, and the dedicated GAPS balloon-borne experiment. 
AMS-02~\cite{AMS02, AMS02foreseen} is expected to be sensitive to a flux
\beq
\phi^{\rm AMS-02}_{\bar d} > 4.5 \cdot 10^{-7} \ ({\rm m}^2\  {\rm sec}\  {\rm sr}\  {\rm GeV}/{\rm n})^{-1} \qquad \left\{ \begin{array}{l} 0.2\ {\rm GeV/n} < T_{\bar d} < 0.8 \ {\rm GeV/n} \\ 2.1\ {\rm GeV/n} < T_{\bar d} < 4.1 \ {\rm GeV/n}  \end{array} \right. .
\eeq
%where the multi-GeV energy region could extend up to possibly 8 GeV thanks to the addition of the RICH detector~\cite{AMS02}. 
GAPS~\cite{GAPS} will use a dedicated technique: an incoming anti-deuteron is slowed down in the detector and forms an exotic atom with one of the nuclei of the detector (replacing a shell electron); eventually the atom de-excites emitting several hard X-rays and the captured antideuteron annihilates on the nucleus producing a characteristic hadronic shower correlated with the X-ray signature.
The Long Duration Balloon (LDB), Ultra Long Duration Balloon (ULDB) and possible satellite-borne (SB) flights foresee a sensitivity respectively to a flux~\cite{GAPS, Profumo2005}
\beq
\left. \begin{array}{l} 
\phi^{\rm GAPS-LBD}_{\bar d} > 1.5 \cdot 10^{-7} \ ({\rm m}^2\  {\rm sec}\  {\rm sr}\  {\rm GeV}/{\rm n})^{-1} \\[2mm]
\phi^{\rm GAPS-ULBD}_{\bar d} > 3.0 \cdot 10^{-8} \ ({\rm m}^2\  {\rm sec}\  {\rm sr}\  {\rm GeV}/{\rm n})^{-1}
 \end{array} \right\}
 \qquad
 0.1\ {\rm GeV/n} < T_{\bar d} < 0.25 \ {\rm GeV/n}
 \eeq
\beq
  \phi^{\rm GAPS-SB}_{\bar d} > 2.6 \cdot 10^{-9} \ ({\rm m}^2\  {\rm sec}\  {\rm sr}\  {\rm GeV}/{\rm n})^{-1}
 \qquad
 0.1\ {\rm GeV/n} < T_{\bar d} < 0.4 \ {\rm GeV/n} 
\eeq

\bigskip

\begin{figure}[t]
\begin{center}
\includegraphics[width=0.8\textwidth]{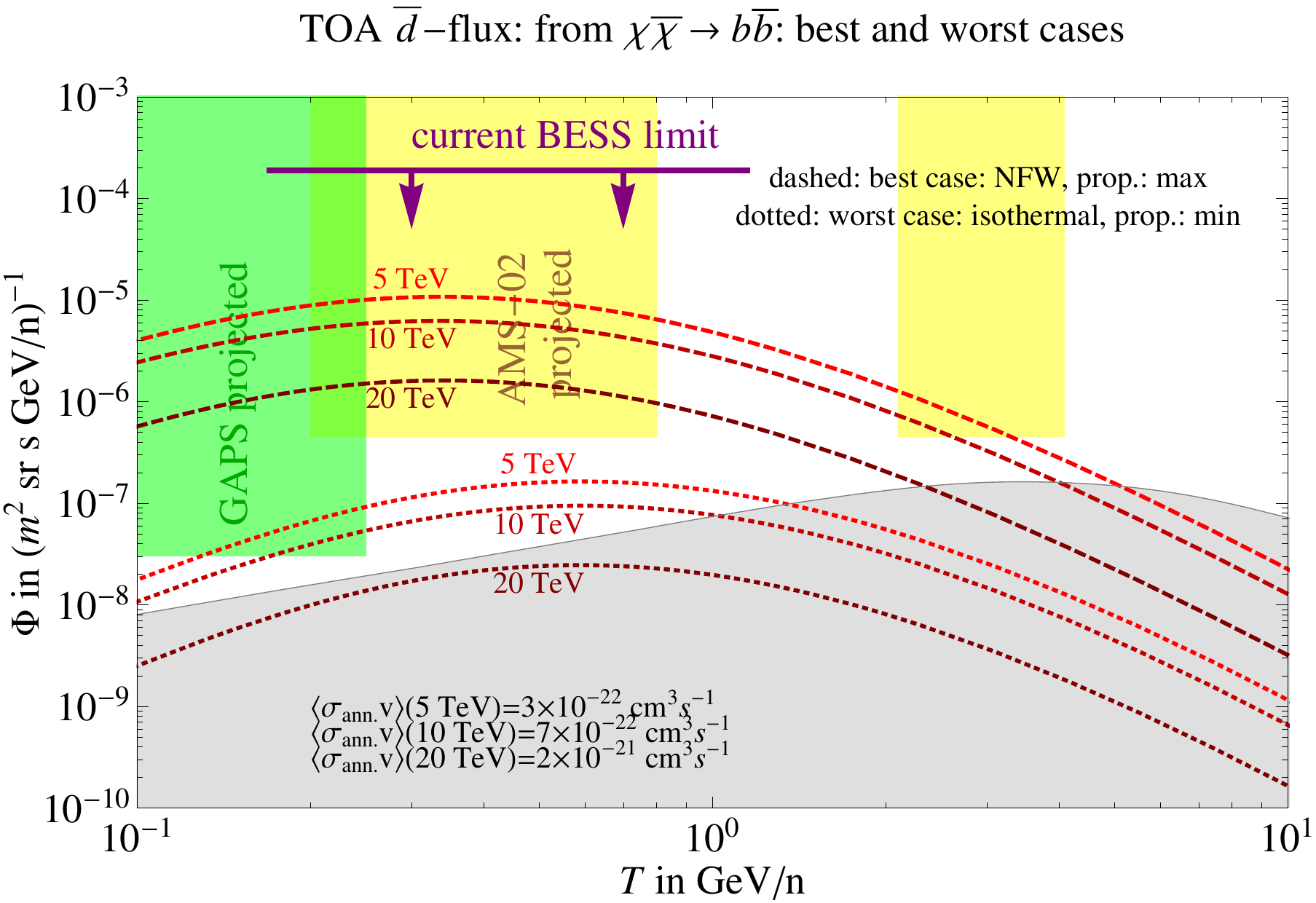}
\caption{\em\label{fig:bestworst} `Best' (DM profile: {\em NFW}, propagation parameters: {\em max}) and `worst' (\,{\em isothermal, min}) scenarios for the detection of heavy dark matter via anti-deuterons. We do not include the 1 TeV case here as it is already excluded by the non-observation of anti-protons. We do not include fluxes for annihilation to $W^+W^-$ either as they are below the background anyway.}
\end{center}
\end{figure}

The final fluxes of antideuterons at Earth, after solar modulation and for the benchmark choice of astrophysical parameters ({\em Einasto, med}) are presented in fig.\fig{final} for the masses, annihilation channels and corresponding annihilation cross sections (table \ref{tab:Msigma}) that we discussed. The fluxes from primary $W^+W^-$ are a few orders of magnitude below the background and also concentrated at high energy, so essentially undetectable. These features can be essentially traced back to the characteristics of the primary fluxes presented in fig.\fig{primary}. The case of Minimal Dark Matter (green line in fig.\fig{final}) falls in this particular category. On the contrary, the fluxes from primary $b\bar{b}$ are large (and relatively flat) at low energies, where the sensitivity of the upcoming experiments is located. Fluxes from larger mass candidates are slightly more suppressed (the source term depends on $M_{\rm DM}^{-2}$, only partly compensated by a larger cross section). It is interesting to note that some of the fluxes can extend into the multi-GeV region (possibly probed by the AMS-02 mission) while remaining above the background.\\ 
It is worth reminding that different choices of the astrophysical parameters lead to order of magnitude changes in the overall normalization, as discussed above: for instance for {\em max} propagation parameters and {\em NFW} profile the predicted fluxes (from $b\bar{b}$) for any DM mass lie well within the AMS-02 area (`best case scenario'), while for {\em min} and {\em isothermal} only the 5 TeV and 10 TeV candidates are predicted to produce a flux that skims through the GAPS region (`worst case scenario'). This is illustrated in fig.\fig{bestworst}.

\section{Conclusions}
\label{conclusions}

We have studied the fluxes of antideuterons produced by heavy Dark Matter annihilations in the galactic halo. We have proceeded in a model independent way, considering a range of DM masses and corresponding large annihilation cross sections motivated by the PAMELA results on the excess in positrons, if interpreted in terms of Dark Matter. These are shown in table~\ref{tab:Msigma}. We have also considered the specific model of Minimal Dark Matter.

We have computed the primary fluxes of $\bar{p}$ and $\bar{n}$ at production via the PYTHIA MonteCarlo code, generated the $\bar{d}$ fluxes (presented in fig.\fig{primary}) via the coalescence analytical model and then propagated them across the galactic halo with the semi-analytic method of the two-zone model. 

Fig.\fig{prop} and fig.\fig{prop2} show the impact of the different choices for the astrophysical propagation parameters, halo DM profile, the coalescence momentum parameter and of the solar modulation effect. The uncertainty on the final predicted fluxes can be large, even of one or two orders of magnitude. 

Fig.\fig{final} shows our main results: the fluxes of antideuterons at Earth for the different masses and primary annihilation channels, together with the predicted experimental sensitivity. For the case of annihilation into a $b\bar{b}$ pair (exemplar of any annihilation into a quark pair) the fluxes are large, concentrated at low energies and therefore in the optimal conditions for detection in the upcoming experiments. Fluxes from the $W^+W^-$ channel (gauge bosons in general) are instead not relevant, and this in particular applies to the model of Minimal Dark Matter that has $W^+W^-$ as the dominant channel.

Our results can be used for predicting the antideuteron fluxes in any other given precise model, where branching ratios for the different primary annihilation channels are known, simply linearly combining and rescaling appropriately the spectra that we presented. 

\bigskip

In summary, we find that if the PAMELA signals in positrons are correctly pointing to DM particles that are heavy (multi-TeV) and with very large annihilation cross sections, then the associated fluxes of antideuterons from the same DM anihilations are very promising for the upcoming experiments (GAPS, AMS-02), if the dominant annihilation channels are into quarks and not gauge bosons (and, of course, not pure leptons). Astrophysical uncertainties have however a sizable impact.

%\begin{figure}[t]
%\begin{center}
%\includegraphics[width=0.4\textwidth]{xxx}\qquad
%\includegraphics[width=0.4\textwidth]{xxx}
%\caption{\em\label{fig:xxx} Caption.}
%\end{center}
%\end{figure}

\paragraph{Acknowledgements}
We thank Michele Papucci for useful discussions. We are particularly indebted to Alessandro Strumia for very inspiring discussions and for constant attention. We are very grateful to Nicolao Fornengo and Fiorenza Donato for carefully reading the manuscript and for useful comments.

The work of C.B.B. is supported by the Bayer Science and Education Foundation and by the Deutsche Akademische Austauschdienst (DAAD).
We also thank the EU Marie Curie Research \& Training network ``UniverseNet" (MRTN-CT-2006-035863) for support.

\bigskip
\appendix

\footnotesize
\begin{multicols}{2}
  
\end{multicols}

\end{document}